\documentclass{article}
\usepackage{amssymb,amsmath,cite,bm}
\def\Op{\mathrm{O}}
\def\spa{\mathrm{span}}
\def\id{\mathbf{1}}
\def\Pe{\mathrm{P}}
\def\Ps{\mathcal{P}}
\def\Vf{\mathbb{V}}
\def\Wf{\mathbb{W}}
\def\Sf{\mathbb{S}}
\def\tr{\mathrm{tr}}
\def\Cf{\mathbb{C}}
\begin{document}

\begin{titlepage}
\begin{center}
{\large \textbf{Classification of matrix product ground states corresponding to one dimensional
chains of two state sites of nearest neighbor interactions}}

\vspace{2\baselineskip}
{\sffamily Amir~H.~Fatollahi\footnote{e-mail: ahfatol@gmail.com},
Mohammad~Khorrami\footnote{e-mail: mamwad@mailaps.org},
Ahmad~Shariati\footnote{e-mail: shariati@mailaps.org}, \&
Amir~Aghamohammadi\footnote{e-mail: mohamadi@alzahra.ac.ir}}

\vspace{2\baselineskip}
{\it Department of Physics, Alzahra University, Tehran 19938-93973, IRAN}
\end{center}
\vspace{2\baselineskip} \textbf{PACS numbers:} 03.65.Fd


\begin{abstract}
\noindent A complete classification is given for one dimensional chains with
nearest neighbor interactions having two states in each site, for which a
matrix product ground state exists. The Hamiltonians and their corresponding
matrix product ground states are explicitly obtained.
\end{abstract}
\end{titlepage}
\newpage
\section{Introduction}
Recently many body problems have absorbed much interest. Among these
problems are quantum spin chain systems. It is desirable to find
the eigenvalues, eigenvectors, and  correlation functions of such
models. But in general it is a very difficult, if not impossible,
task. In fact, there are few models for them the complete spectra
can be obtained in a closed form. The main problem lies in the fact
that the dimension of the Hilbert space for a general many-body
system grows exponentially with the system size, and hence
exponentially many variables are needed to specify the wavefunction
of such a system. Moreover, in some cases although the ground state
is known, its structure turns out to be quite complicated, making
the calculation of correlation functions very difficult. When the
interaction is local, the matrix product state representation
turns out to be one of the techniques which may give us the
opportunity to find the ground state of a many body system.
Recently, extensive studies have been done on the matrix product state formalism
\cite{AKLT,FNW,KSZ,NZ,ASZ,BSZ,VCM,VC,PGVVW}. In this method, first
the ground state is constructed. Then, a Hamiltonian is
written in such a way that the above mentioned state be its
ground state. A matrix product state is a generalization of an
uncorrelated state (the tensor product of one site states).

This method has been used in exactly solvable spin chain models,
spin ladders, and spin systems on two dimensional lattices
\cite{KSZ,NZ,AKS,AKS2,NKZ, VK2}. It can also be applied to
many types of stochastic systems of interacting particles in
one dimensional chains \cite{DVHP,BDer,VK}.
In \cite{AKM}, the matrix product formalism was used to find
the exact ground states of two new spin-1 quantum chains with
nearest neighbor interactions. In \cite{MAs}, the matrix product
states having only spin-flip and parity symmetries, were
classified. It was seen there that there are three distinct
classes of such states. In \cite{SPC}, the matrix product states have
been used to classify quantum phases.

In this article, a classification is presented for the ground states
and their corresponding Hamiltonians, which are obtained by
the matrix product states. This classification applies to models
with nearest neighbor interactions on a one dimensional lattice, for
which the number of states in each site is two. The Heisenberg
spin $1/2$ chain is an example of such models. We classify the models,
for which a matrix product ground state can be obtained, and obtain
the corresponding Hamiltonians and their matrix product ground states.

The scheme of the paper is the following. In section 2, some general
techniques are introduced, mainly to fix the notation.
In section 3 a partial classification is given, based on
possible different Hamiltonians, and without taking into account
that not for all of these Hamiltonians there exists a (nonzero)
matrix product state as the ground state. In section 4 this
classification is completed by finding the corresponding ground states,
and the Hamiltonians and the ground states are explicitly presented.
Section 5 is devoted to the concluding remarks.
\section{General formulation}
To fix the notation, let's recall some basic facts about the
matrix product state. The models addressed here are one dimensional
quantum chains with $N$ sites. The number of possible states of
each site is $d$, and the set $\{e_1,\dots,e_d\}$ is an orthonormal
basis for the Hilbert space corresponding to the states of each site.
Consider a set of $D$ dimensional matrices $\{A^1,\dots,A^d\}$.
Then $\psi$, the (normalized) matrix product state (corresponding
to these matrices) is defined as
\begin{equation}\label{1.1}
\psi:= \frac{1}{\sqrt{\mathcal{Z}}}\,
\tr(A^{\alpha_1}\cdots A^{\alpha_N})\,e_{\alpha_1}\otimes\cdots\otimes e_{\alpha_N}.
\end{equation}
The summation convention is assumed, so that for an index appeared
once and only once as a subscript and once and only once as a
superscript, a summation on that index is assumed.
$\mathcal{Z}$ is the normalization constant:
\begin{equation}\label{1.2}
\mathcal{Z}=\tr(\mathcal{A}^N),
\end{equation}
and
\begin{equation}\label{1.3}
\mathcal{A}:=\delta_{\alpha\,\beta}\,\overline{A^\alpha}\otimes A^\beta,
\end{equation}
where $\bar X$ is the complex conjugate of $X$.

Consider the following equation for the tensor $C$, which is of rank $k$.
\begin{equation}\label{1.4}
C_{\alpha_1\cdots \alpha_k}\,A^{\alpha_1}\cdots A^{\alpha_k}=0.
\end{equation}
The set of all tensors $C$ of rank $k$ satisfying (\ref{1.4}),
is obviously a vector space. Let $\{E^1,\dots\}$ be a basis for
that vector space:
\begin{equation}\label{1.5}
C_{\alpha_1\cdots \alpha_k}=C_a\, E^a_{\alpha_1\cdots \alpha_k}.
\end{equation}
Consider further a family of local Hamiltonians $h$ acting on the
Hilbert space corresponding to $k$ consecutive sites:
\begin{equation}\label{1.6}
h:= \Lambda_{a\,b}\, E^{a\,\dagger}\,E^b,
\end{equation}
where $\Lambda$ is Hermitian and positive semi-definite. These
guarantee that $h$ is hermitian and positive semi-definite, respectively.
In terms of the matrix elements, (\ref{1.6}) can be written as
\begin{equation}\label{1.7}
h^{\alpha_1\cdots\alpha_k}{}_{\beta_1\cdots\beta_k}= \Lambda_{a\,b}\,
(E^{a\,\dagger})^{\alpha_1\cdots\alpha_k}\,E^b_{\beta_1\cdots\beta_k},
\end{equation}
where
\begin{equation}\label{1.8}
h\,(e_{\beta_1}\otimes\cdots\otimes e_{\beta_k})=:
h^{\alpha_1\cdots \alpha_k}{}_{\beta_1\cdots \beta_k}\,e_{\alpha_1}\otimes\cdots\otimes e_{\alpha_k},
\end{equation}
and
\begin{equation}\label{1.9}
E^a=:E^a_{\alpha_1\cdots\alpha_k}\,e^{\alpha_1}\otimes\cdots\otimes e^{\alpha_k},
\end{equation}
and $\{e^1,\dots,e^k\}$ is the basis dual to $\{e_1,\dots,e_k\}$.
Now define the full Hamiltonian $H$ (acting on the Hilbert space corresponding to the
whole lattice) through
\begin{equation}\label{1.10}
H:=\sum_{i=1}^{N-k+1} h_{i, i+k-1},
\end{equation}
where
\begin{equation}\label{1.11}
h_{i,i+k-1}:=\underbrace{\id\otimes\cdots\otimes\id}_{i-1}\otimes h\otimes \underbrace{\id\otimes \cdots \otimes\id}_{N-k-i+1},
\end{equation}
and $\id$ is the identity matrix. Using (\ref{1.4}), or equivalently
\begin{equation}\label{1.12}
E^a_{\alpha_1\cdots\alpha_k}\,A^{\alpha_1}\cdots A^{\alpha_k}=0,
\end{equation}
it is seen that $\psi$, defined through (\ref{1.1}) is an eigenvector of $H$
corresponding to the eigenvalue zero. Also, the fact that $h$ is
positive semi-definite ensures that $H$ is positive semi-definite as
well. So zero is the smallest eigenvalue of $H$, hence $\psi$ is a
ground state corresponding to $H$.

It is seen that the Hamiltonian $H$ constructed through (\ref{1.10}),
describes an interaction in blocks consisting of $k$ consecutive sites.
\section{Hamiltonians, partial classification}
We want to classify models with nearest neighbor interactions ($k=2$), so
(\ref{1.4}) changes to
\begin{equation}\label{1.13}
C_{\alpha\,\beta}\,A^\alpha\,A^\beta=0.
\end{equation}
We also consider cases where the Hilbert space corresponding to each site
is two dimensional ($d=2$). So the set of all the matrices $C$ for which
$A^\alpha$'s are to satisfy (\ref{1.13}), is a vector
subspace $\Vf$ of the $\spa(\Sf)$, where
\begin{equation}\label{1.14}
\Sf:=\{\tau_0, \tau_1, \tau_2, \sigma\},
\end{equation}
and
\begin{align}\label{1.15}
\tau_0&:=\begin{pmatrix}1&0\\0&1\end{pmatrix},\nonumber\\
\tau_1&:=\begin{pmatrix}1&0\\0&-1\end{pmatrix},\nonumber\\
\tau_2&:=\begin{pmatrix}0&1\\1&0\end{pmatrix},\nonumber\\
\sigma&:=\begin{pmatrix}0&1\\-1&0\end{pmatrix}.
\end{align}
Denote the permutation operator by $\Pe$:
\begin{equation}\label{1.16}
(\Pe\,C)_{\alpha\,\beta}:=C_{\beta\,\alpha},
\end{equation}
and the projections to  the eigenspaces of $\Pe$ corresponding to the eigenvalues
$\pm 1$ by $\Pi^\pm$:
\begin{equation}\label{1.17}
\Pi^\pm:=\frac{1}{2}\,[(\id^*\otimes\id^*)\pm\Pe],
\end{equation}
where $\id^*$ is the pullback of identity ($\id$). Defining the symmetric and
antisymmetric projected spaces ($\Vf^+$ and $\Vf^-$
respectively) as
\begin{equation}\label{1.18}
\Vf^\pm:=\Pi^\pm\,\Vf,
\end{equation}
it is seen that $\Vf^\pm$ is a vector subspace of $\spa(\Sf^\pm)$, where
\begin{align}\label{1.19}
\Sf^+&:=\{\tau_0, \tau_1, \tau_2\},\nonumber\\
\Sf^-&:=\{\sigma\}.
\end{align}

The action of an invertible matrix $\Gamma$ on a member of $\spa(\Sf)$ is denoted by
$\Op_\Gamma$:
\begin{equation}\label{1.20}
(\Op_\Gamma\,C)_{\alpha\,\beta}:=\Gamma^\gamma{}_\alpha\,\Gamma^\delta{}_\beta\,
C_{\gamma\,\delta},
\end{equation}
which can be written more compactly as
\begin{equation}\label{1.21}
\Op_\Gamma\,C:=(\Gamma^*\otimes\Gamma^*)\,C,
\end{equation}
or
\begin{equation}\label{1.22}
\Op_\Gamma\,C:=\Gamma^*\,C\,\Gamma,
\end{equation}
where $\Gamma^*$ is the pullback of $\Gamma$:
\begin{equation}\label{1.23}
(\Gamma^*)_\beta{}^\alpha:=\Gamma^\alpha{}_\beta.
\end{equation}
Two spaces $\Vf_1$ and $\Vf_2$ are said to be equivalent to each other, if there is
an invertible matrix $\Gamma$ such that
\begin{equation}\label{1.24}
\Vf_2=\Op_\Gamma\,\Vf_1.
\end{equation}
It is easy to see that the action of $\Gamma$ and the action of any
nonzero multiple of it on a subspace of $\spa(\Sf)$ are the same. So
in order to find all spaces equivalent to $\Vf_1$, it is sufficient
to consider only those $\Gamma$'s which have a unit determinant, that is,
only the matrices belonging to SL$_2(\Cf)$. From now on it is assumed that
the matrices $\Gamma$ acting on subspaces of $\spa(\Cf)$ are members of SL$_2(\Cf)$.

Any member of $\spa(\Sf)$ is characterized by the ordered quartet $(v^0,v^1,v^2,u)$:
\begin{equation}\label{1.25}
C=:v^i\,\tau_i+u\,\sigma.
\end{equation}
Using
\begin{equation}\label{1.26}
[\Gamma^*\otimes\Gamma^*,\Pe]=0,
\end{equation}
it is seen that $\spa(\Sf^+)$ and $\spa(\Sf^-)$ are invariant subspaces of
$\Gamma^*\otimes\Gamma^*$. In fact, as
\begin{equation}\label{1.27}
(\Gamma^*\otimes\Gamma^*)\,\sigma=\det(\Gamma)\,\sigma,
\end{equation}
$\spa(\Sf^-)$ is an eigenspace of $(\Gamma^*\otimes\Gamma^*)$ with eigenvalue one.
This means that under the action of $\Gamma$, $(v^0,v^1,v^2,u)$ is transformed
to $(v'^0,v'^1,v'^2,u)$, that is $u$ remains invariant.

Using
\begin{equation}\label{1.28}
\Gamma^{-1}\,\sigma^{-1}=\sigma^{-1}\,\Gamma^*,
\end{equation}
it is seen that
\begin{equation}\label{1.29}
\sigma^{-1}\,(\Op_\Gamma\,C)\,\sigma^{-1}=\Gamma^{-1}\,(\sigma^{-1}\,C\,\sigma^{-1})\,(\Gamma^*)^{-1},
\end{equation}
so that under the action of $\Gamma$, $(C_1\,\sigma^{-1}\,C_2\,\sigma^{-1})$
is similarity transformed, which means that $\tr(C_1\,\sigma^{-1}\,C_2\,\sigma^{-1})$
remains invariant. It is seen that
\begin{equation}\label{1.30}
\tr(C_1\,\sigma^{-1}\,C_2\,\sigma^{-1})=2\,[-u_1\,u_2-v_1^0\,v_2^0+v_1^1\,v_2^1+v_1^2\,v_2^2].
\end{equation}
As the action of $\Gamma$ on $C$ leaves $u$ invariant, it is seen that
this action leaves the product $(v_1\cdot v_2)$ invariant as well, where
\begin{equation}\label{1.31}
v_1\cdot v_2:=-v_1^0\,v_2^0+v_1^1\,v_2^1+v_1^2\,v_2^2.
\end{equation}

Now consider the action of an SL$_2(\Cf)$ matrix on $\Vf^+$. $\Vf^+$
could be zero-, one-, two-, or three-dimensional. In the first and
last cases, it is invariant under such an action. If $\Vf^+$
is one dimensional, then it either is null (with respect to the
product (\ref{1.31})), or is not null. If it is null, there is a $\Gamma$ such that
\begin{equation}\label{1.32}
\Op_\Gamma\,\Vf^+=\spa\{\tau_0+\tau_1\}.
\end{equation}
If $\Vf^+$ is not null, then there is a $\Gamma$ such that
\begin{equation}\label{1.33}
\Op_\Gamma\,\Vf^+=\spa\{\tau_2\}.
\end{equation}

If $\Vf^+$ is two dimensional, then the subspace of all vectors in
$\spa(\Sf)$ which are normal to it (again with respect to the
product (\ref{1.31})), is one dimensional. This one dimensional
space ($\Wf$), either is null, or is not null. If it is null,
then there is a $\Gamma$ such that
\begin{equation}\label{1.34}
\Op_\Gamma\,\Wf=\spa\{\tau_0+\tau_1\},
\end{equation}
which means that
\begin{equation}\label{1.35}
\Op_\Gamma\,\Vf^+=\spa\{\tau_0+\tau_1,\tau_2\}.
\end{equation}
If $\Wf$ is not null, then there is a $\Gamma$ such that
\begin{equation}\label{1.36}
\Op_\Gamma\,\Wf=\spa\{\tau_1\},
\end{equation}
which means that
\begin{equation}\label{1.37}
\Op_\Gamma\,\Vf^+=\spa\{\tau_0,\tau_2\}.
\end{equation}

To summarize, it is seen that using suitable transformations one can bring
$\Vf^+$ to one of these forms (the final form is denoted by $\Vf^+$ rather
than $\Op_\Gamma\,\Vf^+$):
\begin{align}\label{1.38}
\Vf^+&=\{0\},\\ \label{1.39}
\Vf^+&=\spa\{\tau_2\},\\ \label{1.40}
\Vf^+&=\spa\{\tau_0+\tau_1\},\\ \label{1.41}
\Vf^+&=\spa\{\tau_0,\tau_2\},\\ \label{1.42}
\Vf^+&=\spa\{\tau_0+\tau_1,\tau_2\},\\ \label{1.43}
\Vf^+&=\spa\{\tau_0,\tau_1,\tau_2\}.
\end{align}
In each case, $\Vf$ is either $(\Vf^+\oplus\spa\{\sigma\})$, or a subspace of
$(\Vf^+\oplus\spa\{\sigma\})$ the dimension of which is one less than
the dimension of $(\Vf^+\oplus\spa\{\sigma\})$. One can say that in (\ref{1.25}),
$u$ is either independent of $v$ or a linear function of $v$. In the latter case,
there exists a $w$ such that
\begin{equation}\label{1.44}
u=w\cdot v.
\end{equation}
Each of the cases (\ref{1.38}) to (\ref{1.43}) then can be further analyzed to
obtain nonequivalent possible forms of $\Vf$: Corresponding to (\ref{1.38}),
$u$ is either zero or arbitrary. For the cases (\ref{1.39}) and (\ref{1.40}),
$u$ is either arbitrary or of the form (\ref{1.44}). In the latter case,
\begin{equation}\label{1.45}
u=\mu\,v^2,
\end{equation}
for (\ref{1.39}), and
\begin{equation}\label{1.46}
u=\mu\,v^0,
\end{equation}
for (\ref{1.40}). In this case, one can further make $\mu$ equal to either
zero or one. The reason is that one can take $w$ to be in $\spa(\tau_0-\tau_1)$,
and if it is nonzero there exists an SL$_2(\Cf)$ induced transformation which
leaves $\Vf^+$ invariant and makes $w$ equal to $(\tau_1-\tau_0)/2$.

For (\ref{1.41}), if $u$ is not arbitrary then $w$ can be taken in $\Vf^+$.
If $w$ is not null, it can be transformed to a multiple of $\tau_2$. If $w$
is null, it can be transformed to $(\tau_2-\tau_0)$.

For (\ref{1.42}), if $u$ is not arbitrary then $w$ can be taken in the
$\spa\{\tau_1-\tau_0,\tau_2\}$. The SL$_2(\Cf)$ transformations which leave
$\Vf^+$ invariant are
\begin{align}\label{1.47}
(\tau_0+\tau_1)&\to\frac{1}{\rho}\,(\tau_0+\tau_1),\nonumber\\
(\tau_1-\tau_0)&\to\rho\,(\tau_1-\tau_0)+\lambda\,\tau_2
-\frac{\lambda^2}{2\,\rho}\,(\tau_0+\tau_1),\nonumber\\
\tau_2&\to\pm\left[\tau_2-\frac{\lambda}{2\,\rho}\,(\tau_0+\tau_1)\right].
\end{align}
It shows that $w$ can be transformed to either a constant multiple of $\tau_2$,
or $(\tau_1-\tau_0)/2$.

Finally, for (\ref{1.43}) and if $u$ is not arbitrary, $w$ can be transformed
to either a constant multiple of $\tau_2$, or $(\tau_1-\tau_0)$.

One then arrives at the following nonequivalent forms for $\Vf$.
\begin{align}\label{1.48}
\Vf&=\{0\},\\ \label{1.49}
\Vf&=\spa\{\sigma\},\\ \label{1.50}
\Vf&=\spa\{\tau_2+\mu\,\sigma\},\\ \label{1.51}
\Vf&=\spa\{\tau_2,\sigma\},\\ \label{1.52}
\Vf&=\spa\{\tau_0+\tau_1\},\\ \label{1.53}
\Vf&=\spa\{\tau_0+\tau_1+\sigma\},\\ \label{1.54}
\Vf&=\spa\{\tau_0+\tau_1,\sigma\},\\ \label{1.55}
\Vf&=\spa\{\tau_0,\tau_2+\mu\,\sigma\},\\ \label{1.56}
\Vf&=\spa\{\tau_0+\sigma,\tau_2+\sigma\},\\ \label{1.57}
\Vf&=\spa\{\tau_0+\tau_1,\tau_2+\mu\,\sigma\},\\ \label{1.58}
\Vf&=\spa\{\tau_0+\tau_1+\sigma,\tau_2\},\\ \label{1.59}
\Vf&=\spa\{\tau_0+\tau_1,\tau_2,\sigma\},\\ \label{1.60}
\Vf&=\spa\{\tau_0,\tau_1,\tau_2+\mu\,\sigma\},\\ \label{1.61}
\Vf&=\spa\{\tau_0+\sigma,\tau_1+\sigma,\tau_2\},\\ \label{1.62}
\Vf&=\spa\{\tau_0,\tau_1,\tau_2,\sigma\}.
\end{align}

In each case, one constructs a local Hamiltonian through
(\ref{1.7}), and a ground state for the total Hamiltonian would be
(\ref{1.1}). But this is true only if the matrix product state
(\ref{1.1}) does not vanish. In order to investigate this, and
actually obtain that matrix product state, one has to classify
the solutions for $A^\alpha$'s in (\ref{1.13}), corresponding
to each of the above cases.

Finally, the above classification for the matrices $C$, is a
classification of non equivalent cases. in each case, it is
possible to obtain new $C$ matrices through the action of
an SL$_2(\Cf)$ matrix $\Gamma$. That is, if $\psi$ is a
ground state of the total Hamiltonian the corresponding
local Hamiltonian of which is $h$, then $\psi'$ is a
ground state of a total Hamiltonian the corresponding
local Hamiltonian of which is $h'$, where
\begin{align}\label{1.63}
\psi':=&(\underbrace{\Gamma^{-1}\otimes\cdots\otimes\Gamma^{-1}}_N)\,\psi,\\ \label{1.64}
h':=&(\underbrace{\Gamma^\dagger\otimes\cdots\otimes\Gamma^\dagger}_N)\,h\,
(\underbrace{\Gamma\otimes\cdots\otimes\Gamma}_N).
\end{align}
Note that this (\ref{1.64}) is not a similarity transformation on $h$. So not all
the eigenvectors of $h'$ and $h$ are related to each other through something
like (\ref{1.63}). But the eigenvectors corresponding to the eigenvalue zero
are related to each other this way.
\section{Ground states, full classification}
In this section we want to find the matrices $A^\alpha$
satisfying (\ref{1.13}), corresponding to each of the
nonequivalent spaces of the previous section. Then, corresponding to
each nontrivial pair $(C,A)$ one has a Hamiltonian and
a corresponding ground state. $C$ is nontrivial if it is nonzero. $A$
is nontrivial if the corresponding matrix product state is nonzero.

In general, if one of the metrices $A^\alpha$ is zero, say $A^1$
is zero, then \begin{equation}\label{1.65}
\psi_0:=(e_0)^{\otimes N}
\end{equation}
is a ground state, where $\{e_0,e_1\}$ is an orthonormal basis for the
Hilbert space corresponding to one site. So in all the cases studied below, it is
assumed that non of $A^0$ and $A^1$ vanish, unless otherwise stated.

For the case (\ref{1.48}), $\Vf=\{0\}$, so $C$ is trivial.

Cases (\ref{1.49}) and (\ref{1.50}) can be written as
\begin{equation}\label{1.66}
\nu'\,A^0\,A^1=\nu\,A^1\,A^0,
\end{equation}
where among $\nu$ and $\nu'$ at least one is nonzero. If the ratio of $\nu'$ and $\nu$
is not a root of one, then the trace of any product of $A^0$'s and $A^1$'s containing
both $A^0$ and $A^1$ is zero. Hence $(c_0\,\psi_0+c_1\,\psi_1)$ is a ground state, where
\begin{equation}\label{1.67}
\psi_1:=(e_1)^{\otimes N}.
\end{equation}
This means that both $\psi_0$ and $\psi_1$ are ground states.

If the ratio of $\nu'$ and $\nu$ is a root of one, $M$ is the smallest
positive integer where
\begin{equation}\label{1.68}
\left(\frac{\nu}{\nu'}\right)^M=1,
\end{equation}
and $(N/M)$ is an integer, then another family of ground states is there as well.
In this case, $\psi_k$ is a ground state where
\begin{equation}\label{1.69}
\psi_k:=\sum_{\Ps_{N,k\,M}}\,\zeta(\Ps_{N,k\,M})\,
[e_{\Ps_{N,k\,M}(1)}\otimes\cdots\otimes e_{\Ps_{N,k\,M}(N)}],
\end{equation}
and $\Ps_{N,N'}$ is a function with the domain $\{1,\dots,N\}$, so that
$N'$ of its values are $0$ and $(N-N')$ of its values are $1$.
$\zeta(\Ps_{N,N'})$ is defined as the following. $\Ps_{N,N'}$ corresponds
to a sequence of $0$'s and $1$'s, where the position of the $\ell$'th
$0$ is $i_\ell$. Then
\begin{equation}\label{1.70}
\zeta(\Ps_{N,N'}):=\left(\frac{\nu'}{\nu}\right)^{\sum_\ell(i_\ell-\ell)}.
\end{equation}

The corresponding local Hamiltonian is constructed using
\begin{equation}\label{1.71}
E=\nu'\,e^0\otimes e^1-\nu\,e^1\otimes e^0.
\end{equation}
One has
\begin{equation}\label{1.72}
h=g\,E^\dagger\,E,
\end{equation}
where $g$ is a positive constant. Using the Pauli matrices
\begin{align}\label{1.73}
\sigma_1&:=e_0\,e^1+e_1\,e^0,\nonumber\\
\sigma_3&:=e_0\,e^0-e_1\,e^1,\nonumber\\
\sigma_+&:=e_0\,e^1,\nonumber\\
\sigma_-&:=e_1\,e^0,
\end{align}
this can be rewritten as
\begin{align}\label{1.74}
h&=g\bigg[\frac{|\nu|^2+|\nu'|^2}{4}\,(\id\otimes\id-\sigma_3\otimes\sigma_3)
+\frac{|\nu'|^2-|\nu|^2}{4}\,(\sigma_3\otimes\id-\id\otimes\sigma_3)\nonumber\\
&\qquad-\overline{\nu'}\,\nu\,\sigma_+\otimes\sigma_--\nu'\,\overline{\nu}\,\sigma_-\otimes\sigma_+\bigg].
\end{align}
The full Hamiltonian is the Hamiltonian of a generalized version of the XYZ Heisenberg 
quantum chain, with magnetic field at the boundaries.

In the case (\ref{1.52}), one has
\begin{equation}\label{1.75}
(A^0)^2=0.
\end{equation}
So the trace of any product of $A^0$'s and $A^1$'s vanish if that product contains
two adjacent $A^0$'s. One may guess that any tensor product of $e_0$'s and $e_1$'s
is a ground state, provided that product does not contain two adjacent $e_0$'s.
This is in fact true, and it is easy to see it directly from the Hamiltonian.
The local Hamiltonian corresponding to (\ref{1.52}) is
\begin{equation}\label{1.76}
h=g\,(\id+\sigma_3)\otimes(\id+\sigma_3),
\end{equation}
where $g$ is a positive constant. Obviously the Hamiltonian of the lattice contains
only the operators $(\sigma_3)_i$, which commute with each other. So any
simultaneous eigenvector of these operators is an eigenvector of the Hamiltonian. Noting
\begin{equation}\label{1.77}
\sigma_3\,e_\alpha=(1-2\,\alpha)\,e_\alpha,
\end{equation}
it is seen that $\lambda_{\alpha_1\cdots\alpha_N}$, the eigenvalue of the Hamiltonian
corresponding to the eigenvector $e_{\alpha_1}\otimes\cdots\otimes e_{\alpha_N}$, satisfies
\begin{equation}\label{1.78}
\lambda_{\alpha_1\cdots\alpha_N}=(2-2\,\alpha_1)\,(2-2\,\alpha_2)+\cdots+
(2-2\,\alpha_N),
\end{equation}
which is zero if no two adjacent $\alpha_i$'s are zero, and positive otherwise.

In the case (\ref{1.53}), one has
\begin{equation}\label{1.79}
2\,(A^0)^2+[A^0,A^1]=0.
\end{equation}
Let $\lambda$ be an eigenvalue of $A^0$ and $\Pi_\lambda$ be a projector so that
if $v_{\lambda'}$ is any generalized eigenvector of $A^0$ corresponding to $\lambda'$,
\begin{equation}\label{1.80}
\Pi_\lambda\,v_{\lambda'}=\delta_{\lambda\,\lambda'}\,v_{\lambda'}.
\end{equation}
$\Pi_\lambda$ commutes with $A^0$ and (\ref{1.79}) gives
\begin{equation}\label{1.81}
2\,\Pi_\lambda\,(A^0)^2\,\Pi_\lambda+[A^0,\Pi_\lambda\,A^1\,\Pi_\lambda]=0,
\end{equation}
so
\begin{equation}\label{1.82}
\tr[\Pi_\lambda\,(A^0)^2\,\Pi_\lambda]=0,
\end{equation}
which shows that $\lambda$ should be zero. So $A^0$ is nilpotent. (\ref{1.79})
shows that if $A$ is a linear combination of $A^0$ and $A^1$,
\begin{equation}\label{1.83}
A^0\,A=A'\,A^0,
\end{equation}
where $A'$ is another linear combination of $A^0$ and $A^1$. This shows that if
$B$ is any matrix in the algebra constructed by $A^0$ and $A^1$ and $\id$, and if
$v\in\ker[(A^0)^j]$, then $(B\,v)\in\ker[(A^0)^j]$. Now consider a basis in which
$A^0$ is Jordanian. The vectors in this basis can be grouped into groups $\mathbb{B}_j$,
where $\spa(\mathbb{B}_j)$ is $\mathbb{V}_j$, so that $\mathbb{V}_j$ is a subspace of
$\ker[(A^0)^j]$ but has no nonzero vector in $\ker[(A^0)^{j-1}]$. We want to prove that
\begin{equation}\label{1.84}
\tr(A^0\,B)=0,
\end{equation}
where $B$ is an arbitrary matrix in the algebra constructed by $A^0$ and $A^1$ and $\id$.
To do so, take and arbitrary vector $v$ in $\mathbb{V}_j$. It is seen that
$B\,v$ is in the direct sum of $\mathbb{V}_1$ to $\mathbb{V}_j$. So
$(A^0\,B\,v)$ is in the direct sum of $\mathbb{V}_1$ to $\mathbb{V}_{j-1}$, which
proves (\ref{1.84}). So $\psi_1$ is a ground state.
In this case the local Hamiltonian is like (\ref{1.72}), but with
\begin{equation}\label{1.85}
E=2\,e^0\otimes e^0+e^0\otimes e^1-e^1\otimes e^0,
\end{equation}
so,
\begin{align}\label{1.86}
h&=g\,\bigg[\frac{3}{2}\,\id\otimes\id+\id\otimes\sigma_3+\sigma_3\otimes\id
+\frac{1}{2}\,\sigma_3\otimes\sigma_3\nonumber\\
&\qquad+(1+\sigma_3)\otimes\sigma_1-
\sigma_1\otimes(1+\sigma_3)-\sigma_-\otimes\sigma_+-\sigma_+\otimes\sigma_-\bigg].
\end{align}

For (\ref{1.51}), the trace of any product of $A^0$'s and $A^1$'s containing both $A^0$ and
$A^1$ vanishes. $\psi_0$ and $\psi_1$ are ground states. In this case, $\Vf$ is
two dimensional and a basis for which is $\{E^1,E^2\}$, where
\begin{align}\label{1.87}
E^1&=e^0\otimes e^1,\nonumber\\
E^2&=e^1\otimes e^0.
\end{align}
The local Hamiltonian is then
\begin{align}\label{1.88}
h&=\frac{g_1+g_2}{4}\,(\id\otimes\id-\sigma_3\otimes\sigma_3)+
\frac{g_1-g_2}{4}\,(\sigma_3\otimes\id-\id\otimes\sigma_3)\nonumber\\
&\quad+g_3\,\sigma_+\otimes\sigma_-+\overline{g_3}\,\sigma_-\otimes\sigma_+,
\end{align}
where $g_1$ and $g_2$ are real and nonnegative, and
\begin{equation}\label{1.89}
g_1\,g_2\geq|g_3|^2.
\end{equation}
The full Hamiltonian is a generalized version of the XYZ Heisenberg
quantum chain, with magnetic field at the boundaries.

In the case (\ref{1.54}), $(A^0)^2$ vanishes and $A^0$ and $A^1$ commute with each other.
So (\ref{1.79}) holds and the ground state is $\psi_1$.
In this case, $\Vf$ is two dimensional and a basis for which is $\{E^1,E^2\}$, where
\begin{align}\label{1.90}
E^1&=e^0\otimes e^0,\nonumber\\
E^2&=e^0\otimes e^1-e^1\otimes e^0.
\end{align}
The local Hamiltonian is
\begin{align}\label{1.91}
h&=\frac{g_1+2\,g_2}{4}\,\id\otimes\id+\frac{g_1}{4}\,(\sigma_3\otimes\id+\id\otimes\sigma_3)
\nonumber\\
&\quad+\frac{g_1-2\,g_2}{4}\,\sigma_3\otimes\sigma_3
-g_2\,(\sigma_+\otimes\sigma_-+\sigma_-\otimes\sigma_+)
\nonumber\\
&\quad+\frac{\id+\sigma_3}{2}\otimes(g_3\,\sigma_++\overline{g_3}\,\sigma_-)-
(g_3\,\sigma_++\overline{g_3}\,\sigma_-)\otimes\frac{\id+\sigma_3}{2},
\end{align}
where $g_1$ and $g_2$ are real and nonnegative, and (\ref{1.89}) holds.
The full Hamiltonian corresponds to that of an XXZ quantum spin chain with
constant magnetic field, and an additional boundary interaction.

In the case (\ref{1.55}), one has (\ref{1.66}), with the additional constraint
\begin{equation}\label{1.92}
(A^0)^2+(A^1)^2=0.
\end{equation}
This additional constraint means that ground states other than $\psi_0$ and $\psi_1$ are
possible only if
\begin{equation}\label{1.93}
\left(\frac{\nu}{\nu'}\right)^2=1.
\end{equation}
For $A^0$ and $A^1$ anticommuting, one obtains for even $N$ a ground state
\begin{equation}\label{1.94}
\psi':=\sum_{k=0}^{N/2}(-1)^k\,\sum_{\Ps_{N,2\,k}}\,\zeta(\Ps_{N,2\,k})\,
[e_{\Ps_{N,2\,k}(1)}\otimes\cdots\otimes e_{\Ps_{N,2\,k}(N)}].
\end{equation}
For $A^0$ and $A^1$ commuting, there are two ground states (apart from $\psi_0$ and
$\psi_1$), which are
\begin{equation}\label{1.95}
\psi'_\mathrm{o}:=\sum_{k=1}^{[(N-1)/2]}(-1)^k\,\sum_{\Ps_{N,2\,k+1}}
[e_{\Ps_{N,2\,k+1}(1)}\otimes\cdots\otimes e_{\Ps_{N,2\,k+1}(N)}],
\end{equation}
and
\begin{equation}\label{1.96}
\psi'_\mathrm{e}:=\sum_{k=0}^{[N/2]}(-1)^k\,\sum_{\Ps_{N,2\,k}}
[e_{\Ps_{N,2\,k}(1)}\otimes\cdots\otimes e_{\Ps_{N,2\,k}(N)}].
\end{equation}
In this case, $\Vf$ is two dimensional and a basis for which is $\{E^1,E^2\}$, where
\begin{align}\label{1.97}
E^1&=e^0\otimes e^0+e^1\otimes e^1,\nonumber\\
E^2&=\nu'\,e^0\otimes e^1-\nu\,e^1\otimes e^0.
\end{align}
The local Hamiltonian is
\begin{align}\label{1.98}
h&=\frac{2\,g_1+g_2\,(|\nu'|^2+|\nu|^2)}{4}\,\id\otimes\id+
\frac{2\,g_1-g_2\,(|\nu'|^2+|\nu|^2)}{4}\,\sigma_3\otimes\sigma_3\nonumber\\
&\quad+(g_1-g_2\,\overline{\nu'}\,\nu)\,\sigma_+\otimes\sigma_-
+(g_1-g_2\,\nu'\,\overline{\nu})\,\sigma_-\otimes\sigma_+\nonumber\\
&\quad+g_2\,\frac{|\nu'|^2-|\nu|^2}{4}\,(\sigma_3\otimes\id-\id\otimes\sigma_3)\nonumber\\
&\quad+\frac{g_3}{2}\,[\id\otimes(\nu'\,\sigma_+-\nu\,\sigma_-)
+(\nu'\,\sigma_--\nu\,\sigma_+)\otimes\id\nonumber\\
&\qquad\quad+\sigma_3\otimes(\nu'\,\sigma_++\nu\,\sigma_-)
-(\nu'\,\sigma_-+\nu\,\sigma_+)\otimes\sigma_3]\nonumber\\
&\quad+\frac{\overline{g_3}}{2}\,[\id\otimes(\overline{\nu'}\,\sigma_--\overline{\nu}\,\sigma_+)
+(\overline{\nu'}\,\sigma_+-\overline{\nu}\,\sigma_-)\otimes\id\nonumber\\
&\qquad\quad+\sigma_3\otimes(\overline{\nu'}\,\sigma_-+\overline{\nu}\,\sigma_+)
-(\overline{\nu'}\,\sigma_++\overline{\nu}\,\sigma_-)\otimes\sigma_3],
\end{align}
where $g_1$ and $g_2$ are real and nonnegative, and (\ref{1.89}) holds.

In the case (\ref{1.56}), $(A^0\,A^1)$ vanishes. So the trace of any
product of $A^0$ and $A^1$ containing both $A^0$ and $A^1$ vanishes. But using
\begin{equation}\label{1.99}
(A^0)^2+(A^1)^2+A^0\,A^1-A^1\,A^0=0,
\end{equation}
it is seen that $(A^0)^N$ and $(A^1)^N$ themselves can both be written as
products containing both $A^0$ and $A^1$, with vanishing traces. So no ground state
is obtained this way.

In the case (\ref{1.57}), the trace of any product of $A^0$'s and $A^1$'s containing more than
one $A^0$ vanishes. But as (\ref{1.66}) holds with $\nu'\ne\nu$, the trace of any product
of $A^1$'s and one $A^0$ vanishes as well. So $\psi_1$ is a ground state.
In this case, $\Vf$ is two dimensional and a basis for which is $\{E^1,E^2\}$, where
\begin{align}\label{1.100}
E^1&=e^0\otimes e^0,\nonumber\\
E^2&=\nu'\,e^0\otimes e^1-\nu\,e^1\otimes e^0.
\end{align}
The local Hamiltonian is
\begin{align}\label{1.101}
h&=\frac{2\,g_1+g_2\,(|\nu'|^2+|\nu|^2)}{4}\,\id\otimes\id+
\frac{2\,g_1-g_2\,(|\nu'|^2+|\nu|^2)}{4}\,\sigma_3\otimes\sigma_3\nonumber\\
&\quad-g_2\,(\overline{\nu'}\,\nu\,\sigma_+\otimes\sigma_-
+\nu'\,\overline{\nu}\,\sigma_-\otimes\sigma_+)
+g_2\,\frac{|\nu'|^2-|\nu|^2}{4}\,(\sigma_3\otimes\id-\id\otimes\sigma_3)\nonumber\\
&\quad+\frac{g_3}{2}\,(\nu'\,\id\otimes\sigma_+-\nu\,\sigma_+\otimes\id
+\nu'\,\sigma_3\otimes\sigma_+-\nu\,\sigma_+\otimes\sigma_3)
\nonumber\\
&\quad+\frac{\overline{g_3}}{2}\,(\overline{\nu'}\,\id\otimes\sigma_-
-\overline{\nu}\,\sigma_-\otimes\id
+\overline{\nu'}\,\sigma_3\otimes\sigma_--\overline{\nu}\,\sigma_-\otimes\sigma_3),
\end{align}
where $g_1$ and $g_2$ are real and nonnegative, and (\ref{1.89}) holds.

For the case (\ref{1.58}), (\ref{1.79}) holds meaning that the trace of any product of
$A^0$'s and $A^1$'s other than $(A^1)^N$ vanishes. So $\psi_1$ is a ground state.
In this case, $\Vf$ is two dimensional and a basis for which is $\{E^1,E^2\}$, where
\begin{align}\label{1.102}
E^1&=2\,e^0\otimes e^0+e^0\otimes e^1+e^1\otimes e^0,\nonumber\\
E^2&=e^0\otimes e^1-e^1\otimes e^0.
\end{align}
The local Hamiltonian is
\begin{align}\label{1.103}
h&=\frac{3\,g_1+g_2}{2}\,\id\otimes\id+\frac{g_1-g_2}{2}\,\sigma_3\otimes\sigma_3
+(g_1-g_2)\,(\sigma_+\otimes\sigma_-+\sigma_-\otimes\sigma_+)\nonumber\\
&\quad+g_1\,(\sigma_3\otimes\sigma_1+\sigma_1\otimes\sigma_3)+
g_1\,[(\sigma_3+\sigma_1)\otimes\id+\id\otimes(\sigma_3+\sigma_1)]\nonumber\\
&\quad+\sigma_3\otimes(g_3\,\sigma_-+\overline{g_3}\,\sigma_+)
-(g_3\,\sigma_-+\overline{g_3}\,\sigma_+)\otimes\sigma_3\nonumber\\
&\quad+(\overline{g_3}-g_3)\,(\sigma_+\otimes\sigma_--\sigma_-\otimes\sigma_+)\nonumber\\
&\quad+\id\otimes(g_3\,\sigma_-+\overline{g_3}\,\sigma_+)
-(g_3\,\sigma_-+\overline{g_3}\,\sigma_+)\otimes\id\nonumber\\
&\quad+\frac{\overline{g_3}+g_3}{2}\,(\sigma_3\otimes\id+\id\otimes\sigma_3),
\end{align}
where $g_1$ and $g_2$ are real and nonnegative, and (\ref{1.89}) holds.

In the case (\ref{1.59}), of the quadratic products of $A^0$ and $A^1$ only $(A^1)^2$
is nonvanishing, leading to $\psi_1$ as a ground state.
In this case, $\Vf$ is two dimensional and a basis for which is $\{E^1,E^2\}$, where
\begin{align}\label{1.104}
E^1&=e^0\otimes e^0,\nonumber\\
E^2&=e^0\otimes e^1,\nonumber\\
E^3&=e^1\otimes e^0.
\end{align}
The local Hamiltonian is any hermitian positive semi-definite matrix
so that its kernel includes $e_1\otimes e_1$. This is a nine (real) parameter family.

In the case (\ref{1.60}), $(A^0)^2$ and $(A^1)^2$ vanish and (\ref{1.66}) holds. So the
trace of any product of $A^0$'s and $A^1$'s vanish and no ground state is obtained this way.

The case (\ref{1.61}) is similar to the case (\ref{1.56}) with an additional constraint.
But (\ref{1.56}) leads to vanishing of the trace of any product of $A^0$'s and $A^1$'s.
So for (\ref{1.61}) too, no ground state is obtained this way.

Finally, for (\ref{1.62}) all of the quadratic products of $A^0$ and $A^1$ vanish, so that
no ground state is obtained this way.
\section{Concluding remarks}
To summarize, the local Hamiltonians for which the full Hamiltonian has
a matrix product ground state, and the corresponding matrix product ground state,
are the following.
\begin{align}\label{1.105}
h&=g\bigg[\frac{|\nu|^2+|\nu'|^2}{4}\,(\id\otimes\id-\sigma_3\otimes\sigma_3)
+\frac{|\nu'|^2-|\nu|^2}{4}\,(\sigma_3\otimes\id-\id\otimes\sigma_3)\nonumber\\
&\qquad-\overline{\nu'}\,\nu\,\sigma_+\otimes\sigma_--\nu'\,\overline{\nu}\,\sigma_-\otimes\sigma_+\bigg],
\end{align}
with $\psi_0$ and $\psi_1$ being ground states (of the full Hamiltonian).
If $(\nu'/\nu)$ is a root of $1$, then $\psi_k$'s are also ground states, where
\begin{equation}\label{1.106}
\psi_k:=\sum_{\Ps_{N,k\,M}}\,\zeta(\Ps_{N,k\,M})\,
[e_{\Ps_{N,k\,M}(1)}\otimes\cdots\otimes e_{\Ps_{N,k\,M}(N)}],
\end{equation}
and $M$ is the smallest positive integer so that $(\nu'/\nu)^M$ is $1$.

\begin{equation}\label{1.107}
h=g\,(\id+\sigma_3)\otimes(\id+\sigma_3),
\end{equation}
with any tensor product of $e_0$'s and $e_1$'s, not containing
two adjacent $e_0$'s being ground states.

\begin{align}\label{1.108}
h&=g\,\bigg[\frac{3}{2}\,\id\otimes\id+\id\otimes\sigma_3+\sigma_3\otimes\id
+\frac{1}{2}\,\sigma_3\otimes\sigma_3\nonumber\\
&\qquad+(1+\sigma_3)\otimes\sigma_1-
\sigma_1\otimes(1+\sigma_3)-\sigma_-\otimes\sigma_+-\sigma_+\otimes\sigma_-\bigg],
\end{align}
with $\psi_1$ being a ground state.

\begin{align}\label{1.109}
h&=\frac{g_1+g_2}{4}\,(\id\otimes\id-\sigma_3\otimes\sigma_3)+
\frac{g_1-g_2}{4}\,(\sigma_3\otimes\id-\id\otimes\sigma_3)\nonumber\\
&\quad+g_3\,\sigma_+\otimes\sigma_-+\overline{g_3}\,\sigma_-\otimes\sigma_+,
\end{align}
where $g_1$ and $g_2$ are real and nonnegative, and
\begin{equation}\label{1.110}
g_1\,g_2\geq|g_3|^2.
\end{equation}
Here $\psi_0$ and $\psi_1$ are ground states.

\begin{align}\label{1.111}
h&=\frac{g_1+2\,g_2}{4}\,\id\otimes\id+\frac{g_1}{4}\,(\sigma_3\otimes\id+\id\otimes\sigma_3)
\nonumber\\
&\quad+\frac{g_1-2\,g_2}{4}\,\sigma_3\otimes\sigma_3
-g_2\,(\sigma_+\otimes\sigma_-+\sigma_-\otimes\sigma_+)
\nonumber\\
&\quad+\frac{\id+\sigma_3}{2}\otimes(g_3\,\sigma_++\overline{g_3}\,\sigma_-)-
(g_3\,\sigma_++\overline{g_3}\,\sigma_-)\otimes\frac{\id+\sigma_3}{2},
\end{align}
where $g_1$ and $g_2$ are real and nonnegative, and (\ref{1.110}) holds.
$\psi_1$ is a ground state.

\begin{align}\label{1.112}
h&=\frac{2\,g_1+g_2\,(|\nu'|^2+|\nu|^2)}{4}\,\id\otimes\id+
\frac{2\,g_1-g_2\,(|\nu'|^2+|\nu|^2)}{4}\,\sigma_3\otimes\sigma_3\nonumber\\
&\quad+(g_1-g_2\,\overline{\nu'}\,\nu)\,\sigma_+\otimes\sigma_-
+(g_1-g_2\,\nu'\,\overline{\nu})\,\sigma_-\otimes\sigma_+\nonumber\\
&\quad+g_2\,\frac{|\nu'|^2-|\nu|^2}{4}\,(\sigma_3\otimes\id-\id\otimes\sigma_3)\nonumber\\
&\quad+\frac{g_3}{2}\,[\id\otimes(\nu'\,\sigma_+-\nu\,\sigma_-)
+(\nu'\,\sigma_--\nu\,\sigma_+)\otimes\id\nonumber\\
&\qquad\quad+\sigma_3\otimes(\nu'\,\sigma_++\nu\,\sigma_-)
-(\nu'\,\sigma_-+\nu\,\sigma_+)\otimes\sigma_3]\nonumber\\
&\quad+\frac{\overline{g_3}}{2}\,[\id\otimes(\overline{\nu'}\,\sigma_--\overline{\nu}\,\sigma_+)
+(\overline{\nu'}\,\sigma_+-\overline{\nu}\,\sigma_-)\otimes\id\nonumber\\
&\qquad\quad+\sigma_3\otimes(\overline{\nu'}\,\sigma_-+\overline{\nu}\,\sigma_+)
-(\overline{\nu'}\,\sigma_++\overline{\nu}\,\sigma_-)\otimes\sigma_3],
\end{align}
where $g_1$ and $g_2$ are real and nonnegative, and (\ref{1.110}) holds.
$\psi_0$ and $\psi_1$ are ground states. If $\nu'=-\nu$, then $\psi'$ is
a ground state as well, where
\begin{equation}\label{1.113}
\psi':=\sum_{k=0}^{N/2}(-1)^k\,\sum_{\Ps_{N,2\,k}}\,\zeta(\Ps_{N,2\,k})\,
[e_{\Ps_{N,2\,k}(1)}\otimes\cdots\otimes e_{\Ps_{N,2\,k}(N)}].
\end{equation}
If $\nu'=\nu$, then $\psi'_\mathrm{o}$ and $\psi'_\mathrm{e}$ are ground states
as well (apart from $\psi_0$ and $\psi_1$), where
\begin{equation}\label{1.114}
\psi'_\mathrm{o}:=\sum_{k=1}^{[(N-1)/2]}(-1)^k\,\sum_{\Ps_{N,2\,k+1}}
[e_{\Ps_{N,2\,k+1}(1)}\otimes\cdots\otimes e_{\Ps_{N,2\,k+1}(N)}],
\end{equation}
and
\begin{equation}\label{1.115}
\psi'_\mathrm{e}:=\sum_{k=0}^{[N/2]}(-1)^k\,\sum_{\Ps_{N,2\,k}}
[e_{\Ps_{N,2\,k}(1)}\otimes\cdots\otimes e_{\Ps_{N,2\,k}(N)}].
\end{equation}

\begin{align}\label{1.116}
h&=\frac{2\,g_1+g_2\,(|\nu'|^2+|\nu|^2)}{4}\,\id\otimes\id+
\frac{2\,g_1-g_2\,(|\nu'|^2+|\nu|^2)}{4}\,\sigma_3\otimes\sigma_3\nonumber\\
&\quad-g_2\,(\overline{\nu'}\,\nu\,\sigma_+\otimes\sigma_-
+\nu'\,\overline{\nu}\,\sigma_-\otimes\sigma_+)
+g_2\,\frac{|\nu'|^2-|\nu|^2}{4}\,(\sigma_3\otimes\id-\id\otimes\sigma_3)\nonumber\\
&\quad+\frac{g_3}{2}\,(\nu'\,\id\otimes\sigma_+-\nu\,\sigma_+\otimes\id
+\nu'\,\sigma_3\otimes\sigma_+-\nu\,\sigma_+\otimes\sigma_3)
\nonumber\\
&\quad+\frac{\overline{g_3}}{2}\,(\overline{\nu'}\,\id\otimes\sigma_-
-\overline{\nu}\,\sigma_-\otimes\id
+\overline{\nu'}\,\sigma_3\otimes\sigma_--\overline{\nu}\,\sigma_-\otimes\sigma_3),
\end{align}
where $g_1$ and $g_2$ are real and nonnegative, and (\ref{1.110}) holds.
$\psi_1$ is a ground state.

\begin{align}\label{1.117}
h&=\frac{3\,g_1+g_2}{2}\,\id\otimes\id+\frac{g_1-g_2}{2}\,\sigma_3\otimes\sigma_3
+(g_1-g_2)\,(\sigma_+\otimes\sigma_-+\sigma_-\otimes\sigma_+)\nonumber\\
&\quad+g_1\,(\sigma_3\otimes\sigma_1+\sigma_1\otimes\sigma_3)+
g_1\,[(\sigma_3+\sigma_1)\otimes\id+\id\otimes(\sigma_3+\sigma_1)]\nonumber\\
&\quad+\sigma_3\otimes(g_3\,\sigma_-+\overline{g_3}\,\sigma_+)
-(g_3\,\sigma_-+\overline{g_3}\,\sigma_+)\otimes\sigma_3\nonumber\\
&\quad+(\overline{g_3}-g_3)\,(\sigma_+\otimes\sigma_--\sigma_-\otimes\sigma_+)\nonumber\\
&\quad+\id\otimes(g_3\,\sigma_-+\overline{g_3}\,\sigma_+)
-(g_3\,\sigma_-+\overline{g_3}\,\sigma_+)\otimes\id\nonumber\\
&\quad+\frac{\overline{g_3}+g_3}{2}\,(\sigma_3\otimes\id+\id\otimes\sigma_3),
\end{align}
where $g_1$ and $g_2$ are real and nonnegative, and (\ref{1.110}) holds.
$\psi_1$ is a ground state.

Finally, for any local Hamiltonian the kernel of which includes $e_0\otimes e_0$,
a ground state of the full Hamiltonian is $\psi_1$.

All Hamiltoninas related to the above through (\ref{1.64}), have corresponding
ground states constructed through (\ref{1.63}).
\\[\baselineskip]
\textbf{Acknowledgement}:  This work was partially
supported by the research council of the Alzahra University.

\end{document}